\documentclass[11pt, preprint]{article}
\usepackage{color} \usepackage{cancel}
\usepackage{float} \usepackage{graphicx}
\usepackage{subfigure} \usepackage{caption}	
\usepackage{dcolumn}
\usepackage[toc,page]{appendix} \usepackage{authblk}	 		
\usepackage{indentfirst}	
\usepackage{authblk}	 		
\usepackage{multirow}     
\usepackage{hyperref}     
\usepackage{amsfonts, amssymb, amsmath}
\usepackage[left=2.5cm,right=2.5cm,top=2.9cm,bottom=2.9cm,a4paper]{geometry}
\usepackage[noadjust]{cite}
\usepackage{filecontents}
\usepackage{soul}
\usepackage{xcolor}
\setstcolor{red}

\newcommand{\be}{\begin{equation}}
\newcommand{\ee}{\end{equation}}
\newcommand{\bea}{\begin{eqnarray}}
\newcommand{\eea}{\end{eqnarray}}

\graphicspath{{./figures/}}
\hypersetup{ colorlinks   = true, 
urlcolor     = blue, 
linkcolor    = blue, 
citecolor    = red 	 
}

\title{\Large\bf Modified inflationary models based on scalar-torsion gravity}
\author[1]{I. V. Fomin\thanks{ingvor@inbox.ru}}
\author[1,2,3]{S. V. Chervon\thanks{chervon.sergey@gmail.com}}
\author[2]{K. A. Bolshakova\thanks{bolshakova.ktrn@gmail.com}}
\affil[1]{\small \it  Bauman Moscow State Technical University, Russia}
\affil[2]{\small \it Ulyanovsk State Pedagogical University, Ulyanovsk, Russia}
\affil[3]{\small \it Kazan Federal University, Kazan, Russia}

\begin{document}

\maketitle
\begin{abstract}
In this work, we consider the corrections to the cosmological models based on the teleparralel equivalent of general relativity and the scalar-torsion gravity implying non-minimal coupling between scalar field and torsion.
To determine these corrections, we consider a power-law parameterization of the deviations between teleparralel equivalent of general relativity and the scalar-torsion gravity.
The impact of these deviations on cosmological dynamics, scalar field potential and parameters of cosmological perturbations is considered for different inflationary models.

\end{abstract}

\section{Introduction}

In modern cosmological models, an important role is played by the description of two stages of the accelerated expansion of the universe, namely the primary inflation of the early universe \cite{Baumann:2014nda,Chervon:2019sey} and the second inflation in the modern era \cite{Steinhardt:2008nk} of the evolution of the universe.

To describe the cosmological dynamics of the universe at the different stages of its evolution and observed large-scale structure of the universe, general relativity \cite{Baumann:2014nda,Chervon:2019sey} and various modified theories of gravity are being considered \cite{Clifton:2011jh,Nojiri:2017ncd}. One such approach is the use of the teleparallel equivalent of general relativity (TEGR), in which the scalar curvature $R$ is replaced by torsion scalar $T$ \cite{deAndrade:2000kr,Arcos:2004tzt} and its different modifications \cite{Awad:2017ign,Krssak:2018ywd,Gonzalez-Espinoza:2019ajd,Gonzalez-Espinoza:2020azh,Gonzalez-Espinoza:2020jss} as well.

Equations of cosmological dynamics for the case of $f(T)$-gravity of the second order as in the case of general relativity, however, instead of using the curvature defined through the Levi-Civita connection, in the case of $f(T)$-gravity the Weitzenb\"ock connection is used, which implies zero curvature and non-zero torsion \cite{Cai:2015emx}. The fact that the dynamic equations for $f(T)$-gravity are second order makes these theories simpler than those modified by curvature invariants, and a deeper study of this type of model is of considerable interest for theoretical cosmology.

It should also be noted that the speed of gravitational waves for the case of $f(T)$-gravity is equal to the speed of light in vacuum, which is similar to both the case of general relativity and scalar-tensor theories of gravity and satisfies the restrictions on the speed of gravitational waves from astrophysical objects according to modern observations \cite{Li:2018ixg}.

Thus, the teleparallel equivalent of Einstein gravity and its modifications are the basis for the construction of cosmological models verifiable by observational data, which corresponds to the growing interest in this type of gravity models and their physical applications in modern cosmology.

In ~\cite{Fomin:2020woj,Fomin:2020caa,Fomin:2020hdf,Fomin:2022ozv} methods were proposed for analyzing cosmological models with these modified theories of gravity based on functional and parametric connections to General Relativity, which made it possible to directly assess the influence of the modifications of Einstein gravity on the background dynamics and parameters of cosmological perturbations. In ~\cite{Chervon:2023gio} this approach was used for reconstruction of scalar-torsion gravity theories from the physical potentials of a scalar field.

In these work we present the generalization of this approach to the case of TEGR modifications implying non-minimal coupling between a scalar field and torsion. In the context of the proposed approach to this interpretation of the description of the inflationary scenario, we consider the influence of the non-minimal coupling of the scalar field and torsion on the nature of cosmological dynamics, the scalar field potential and the parameters of cosmological perturbations.

The article is organized as follows. Section \ref{Section2} discusses the equations of background dynamics for the case of scalar-torsional gravity. Section \ref{Section3} discusses the application of the power-law parameterization method to the modifications of the teleparallel equivalent of general relativity on the parameters of inflationary models. Section \ref{Section4} discusses examples of cosmological inflation models based on the proposed approach. It is shown that verification of inflationary models by observational restrictions corresponds only to the assessment of parameter values in the power-law parameterization of the influence of TEGR modifications. Section \ref{Section5} provides an overview of the main results of this work.

\section{Dynamical equations for  cosmological models based on the scalar-torsion gravity}\label{Section2}

The action of generalized scalar-torsion theory with self-interaction of Galilean-like field in the system of units $M^2_{pl}=(8\pi G)^{-1}=1$ can be defined as follows \cite{Gonzalez-Espinoza:2019ajd}
	\begin{eqnarray}
		S=\int d^4x e \left[\frac{1}{2}F(\phi)T+P(\phi,X)-\square \phi G(\phi,X)\right],
		\label{act}
	\end{eqnarray}
where $P$, $F$, $G$ are functions of the scalar field  $\phi$ and its kinetic energy $X=-\frac{\left(\partial\phi\right)^2}{2}$, $T$ is the torsion scalar, $e$ are the tetrad components in a coordinate base $e=det(e^A_{\mu})=\sqrt{-g}$ and $G(\phi,X)$ is the Galileon-type field self-interaction.
	
We will consider a special case with following functions
	\begin{equation}
		F=F(\phi)\neq 0, G=0.
	\end{equation}
	
Thus, for this case, the action  Eq.~(\ref{act}) is reduced to the following form
	\begin{equation}
		S=\int  d^4x e \left[\frac{1}{2}F(\phi)T+P(\phi,X) \right].
		\label{act1}
	\end{equation}
	
Now, we consider the tetrad field
	\begin{equation}
		e^A_\mu=diag\left\lbrace 1,a,a,a \right\rbrace,
		\label{e}
	\end{equation}
corresponding to the homogeneous and isotropic Friedman-Robertson-Walker (FRW) universe with metric
	\begin{equation}
		ds^2=-dt^2+a^2\delta_{ij}dx^idx^j,
	\end{equation}
where $a=a(t)$ is the scale factor which dependent on cosmic time $t$ for expanding universe.


Thus, the background equations for the action Eq.~(\ref{act1}) and tetrad field (\ref{e}) are \cite{Gonzalez-Espinoza:2019ajd,Gonzalez-Espinoza:2020azh,Gonzalez-Espinoza:2020jss}
	\begin{equation}
		P-2XP_{,X}-3 H^2F=0,
		\label{1}
	\end{equation}
	\begin{equation}
		P-3 H^2F-2\dot{H}F-2\dot{F}H=0,
		\label{2}
	\end{equation}
	\begin{equation}
		P_{,\phi}-2XP_{,\phi X}-3 H^2_{,\phi}-3HP_{, X}\dot{\phi}-\left[ P_{, X}+2P_{, XX}\right]. \ddot{\phi}=0.
		\label{3}
	\end{equation}
	
Let's consider a particular case where the function $P$ is defined as follows
	\begin{equation}
		P=-\omega (\phi) X+V(\phi)=-\frac{\omega (\phi)}{2}\dot{\phi}^2+V(\phi),
	\end{equation}
	here $\omega (\phi)$ is some function.
	
Then the system of equations takes the form Eq.~(\ref{1})-(\ref{2}) we obtain
	\begin{equation}
		\omega (\phi)\dot{\phi}^2 =-2H\dot{F} - 2F\dot{H},
		\label{pole}
	\end{equation}
	\begin{equation}
		V(\phi)=3H^2F+2\dot{H}F+2\dot{F}H,
		\label{poten}
	\end{equation}

If $\omega (\phi)=1$, we obtain the expressions described in \cite{Gonzalez-Espinoza:2019ajd} and these Eq.~(\ref{pole})-(\ref{poten}) are similar to those derived in the paper \cite{Chervon:2023gio}.
	
\section{Inflationary models with connection $F\sim H^n$}\label{Section3}

Now, we consider the power-law connection between coupling function $F=F(\phi)$ and the Hubble parameter $H=H(t)$ to characterize the influence of the non-minimal coupling between the scalar filed and torsion on the parameters of cosmological models under consideration.
	
We define this connection as follows \cite{Chervon:2023gio}
\begin{equation}
\label{PLPM}
F(\phi(t))=\left(\frac{H(t)}{\lambda}\right)^n,
\end{equation}
where $\lambda$ is an positive constant and $n$ is the parameter which defines the influence of the non-minimal coupling between the scalar filed and torsion, since for $n=0$ one has the case $F=1$ corresponding to minimal coupling or TEGR.

Dynamic equations(\ref{pole})-(\ref{poten}), in these case, have the following form
\begin{equation}
V(\phi)=3\lambda^{-n}H^{2+n}+\lambda^{-n}\dot{H}H^n(1+n),
\label{poten_3}
\end{equation}
\begin{equation}
\omega\dot{\phi}^2=-2\lambda^{-n}\dot{H}H^n(n+1).
\label{pole_3}
\end{equation}

In this work, unlike paper \cite{Chervon:2023gio}, where kinetic function was considered as $\omega=\omega(\phi)$, the function $\omega=\pm1$ corresponding to the canonical (for $\omega=1$) and phantom (for $\omega=-1$) scalar fields.

Thus, equations (\ref{poten_3})--(\ref{pole_3}) can be rewritten as
\begin{equation}
V(\phi)=\frac{H^{n}}{\lambda^{n}}\left(3H^2+\dot{H}(1+n)\right) %
\label{poten_4a}
\end{equation}
\begin{equation}
\omega\dot{\phi}^2=-2\left(\frac{H}{\lambda}\right)^{n}\dot{H}(n+1)
=-\frac{2}{\lambda^{n}}\frac{d}{dt}\left(H^{n+1}\right). %
\label{pole_4}
\end{equation}
	
From Eq. (\ref{pole_4}) we obtain
	\begin{equation}
		H(t)=\lambda\left[C-\frac{\omega}{2\lambda}\int\dot{\phi}^2dt\right]^{\frac{1}{1+n}},
		\label{S1}
	\end{equation}
where $C$ is the constant of integration.

Thus, the non-minimal connection between the field and torsion induces corrections to the dynamics of cosmological expansion for the TEGR case. In these models, such corrections are expressed through parameter $n$.

For case $n=0$, cosmological models are reduced to ones based on TEGR
	\begin{equation}
		V_{(TEGR)}=3H^2_{(TEGR)}+\dot{H}_{(TEGR)}, %
		\label{TEGR1}
	\end{equation}
	\begin{equation}
		\omega\dot{\phi}^2_{(TEGR)}=-2\dot{H}_{(TEGR)}. %
		\label{TEGR2}
	\end{equation}

From Eq. (\ref{TEGR2}) we get following expression for the Hubble partameter
	\begin{equation}
		H_{(TEGR)}(t)=C-\frac{\omega}{2}\int\dot{\phi}^2_{(TEGR)}dt.
		\label{TEGR3}
	\end{equation}

For the case $n=-1$ one has de Sitter solutions $\phi=const$, $V=const$ and $H=const$.

\subsection{Modified background parameters}
A non-zero constant $C\neq0$ indicates the presence of a cosmological constant in the model, since, in this case, potential (\ref{poten_3}) can be expressed as $V(\phi)=\Lambda+\tilde{V}(\phi)$.
On the contrary, condition $C=0$ means the absence of a cosmological constant in the model $\Lambda=0$.

Now, we introduce notation $Y=-\frac{\omega}{2\lambda}\int\dot{\phi}^2dt$ and rewrite expression (\ref{S1}) as follows
	\begin{equation}
		H(t)=\lambda\left(C+Y(t)\right)^{\frac{1}{1+n}}.
		\label{S2H}
	\end{equation}

At the inflationary quasi de Sitter stage of the evolution of the universe, conditions $H\simeq const$ and $Y\ll C$ are satisfied, therefore
	\begin{equation}
		H(t)=\lambda C^{\frac{1}{1+n}}+\frac{\lambda C^{-\frac{n}{1+n}}}{(1+n)}Y+{\mathcal O}\left(Y^{2}\right).
		\label{S3H}
	\end{equation}

Thus, we can write the Hubble parameter at a first order as
	\begin{equation}
H(t)\simeq \tilde{C}-\frac{\tilde{\omega}}{2}\int\dot{\phi}^2dt,
		\label{S4}
	\end{equation}
where $\tilde{C}=\lambda C^{\frac{1}{1+n}}$ and $\tilde{\omega}=\frac{\omega C^{-\frac{n}{1+n}}}{(1+n)}$.

Thus, at the inflationary stage, for the case of a non-zero constant $C\neq0$, the non-minimal coupling between the scalar field and torsion does not change the type of accelerated expansion of the early universe for the same scalar field evolution $\phi(t)=\phi(t)_{(TEGR)}$.

However, for the case of a non-zero constant $C=0$, such a coupling can significantly affect the type of accelerated expansion of the early universe.

Also, under condition $H\simeq const$ from (\ref{poten_3}) and (\ref{TEGR1}) we have
	\begin{equation}
V(\phi)\simeq F(\phi)V_{(TEGR)}(\phi).
		\label{S5}
	\end{equation}

Thus, the non-minimal coupling between the scalar field and torsion can have a significant impact on the potential of the scalar field, which is determined by the non-minimal coupling function $F$.

\subsection{Modified parameters of cosmological perturbations}

From the standpoint of the inflationary paradigm, the origin of the large-scale structure of the universe is due to quantum fluctuations of the scalar field and corresponding perturbations of the space-time metric at the inflationary stage of the evolution of the universe. Also, the theory of cosmological perturbations predicts the presence of relict gravitational waves \cite{Baumann:2014nda,Chervon:2019sey}.

The influence of cosmological perturbations on the anisotropy and polarization of CMB makes it possible to determine the correctness of inflationary models based on observational restrictions on the values of the parameters of cosmological perturbations.

According to modern observational data \cite{Planck:2018vyg,Tristram:2021tvh}, constraints on the parameters of cosmological perturbations are
	\begin{equation}
P_{S}=2.1\times10^{-9},~~~~n_{S}=0.9663\pm 0.0041,~~~~r=\frac{P_{T}}{P_{S}}<0.032,
		\label{PARCONSTRAINTS}
	\end{equation}
where $P_{S}$ and $P_{T}$ are the power spectra of scalar and tensor perturbations correspondingly, $n_{S}$ is the spectral index of scalar perturbations and $r$ is the tensor-to-scalar ratio.

For models of cosmological inflation based on scalar-torsion gravity, cosmological perturbations were previously considered in various works (see, for example, in \cite{Gonzalez-Espinoza:2019ajd,Gonzalez-Espinoza:2020azh,Gonzalez-Espinoza:2020jss}).

For the power-law connection between the non-minimal coupling function and the Hubble parameter $F\sim H^{n}$, the parameters of cosmological perturbations were obtained in \cite{Chervon:2023gio} as
\begin{eqnarray}
\label{PERTG2}
&&{\mathcal P}_{S}=\frac{\lambda^{n}}{2(n+1)\epsilon}\left(\frac{H}{2\pi}\right)^{2},\\
\label{PERTG3}
&&n_{S}-1=(n-4)\epsilon+2\delta,\\
\label{PERTG4}
&&r=\frac{{\mathcal P}_{T}}{{\mathcal P}_{S}}=16(n+1)\epsilon,
\end{eqnarray}
where the slow-roll parameters are defined as follow
\begin{eqnarray}
\label{SR1}
\epsilon=-\frac{\dot{H}}{H^{2}}\ll1,
\label{SR2}
&&\delta=-\frac{\ddot{H}}{2H\dot{H}}\ll1.
\end{eqnarray}

As one can see from expressions (\ref{PERTG2})--(\ref{PERTG4}), when taking into account the non-minimal coupling between the scalar field and torsion, an arbitrary inflationary model can correspond to observational constraints(\ref{PARCONSTRAINTS}).

Therefore, the procedure for verifying cosmological inflation models, in this case, can be reduced to determining the possible values of parameter $n$ in the power-law parameterization of TEGR modifications (\ref{PLPM}).

\section{Modified inflationary models}\label{Section4}

Now, let us consider the influence of the non-minimal coupling between the scalar field and torsion on the parameters of inflationary models.

In this section we will consider two types of solutions (\ref{S1}) for the cases $C=0$ and $C\neq0$ for power-law and exponential power-law inflation correspondingly.

It is obvious that the proposed approach can be considered for arbitrary models of cosmological inflation, and these cases are chosen as a clear illustration of the analysis of cosmological models based on the power-law parameterization of the influence of TEGR modifications on the model's parameters.

\subsection{\label{sec:level4}Modified power-law inflation}
	
Firsyly, we will consider corrections to standard power-law inflation \cite{Chervon:2019sey} induced by a non-minimal coupling between the scalar field and torsion.

For this purpose we consider the following canonical scalar field
	\begin{equation}
		\phi(t)=A\ln(Bt),
		\label{S2}
	\end{equation}
where $A$ and $B$ are the constants.

From Eq. (\ref{S1}) for $\omega=1$ and $C=0$ we obtain
	\begin{equation}
		H(t)=\left(\frac{m}{t}\right)^{\frac{1}{1+n}},~~~~m=\frac{\lambda^{n}A^{2}}{2}.
		\label{S3}
	\end{equation}

Corresponding scale factor is
	\begin{equation}
a(t)\propto\exp\left[\left(1+\frac{1}{n}\right)t\left(\frac{m}{t}\right)^{\frac{1}{1+n}}\right],~~~~~n\neq0.
		\label{S4A}
	\end{equation}

Thus, we obtain intermediate inflation for scalar-torsion gravity instead of power-law inflation for TEGR.

From inverse dependence $t=t(\phi)$, connection between coupling function and the Hubble parameter $F=\left(H/\lambda\right)^{n}$ and Eq. (\ref{poten_4a}) we obtain
	\begin{equation}
		V(\phi)=\frac{3}{\lambda^{n}}(mB)^{\frac{n+2}{n+1}}\exp\left(-\frac{(n+2)\phi}{(n+1)A}\right)
-\frac{mB^{2}}{\lambda^{n}}\exp\left(-2\frac{\phi}{A}\right),
		\label{Sv}
	\end{equation}
	\begin{equation}
		F(\phi)=\left(\frac{mB}{\lambda}\right)^{\frac{n}{n+1}}\exp\left(-\frac{n\phi}{A(1+n)}\right).
		\label{Sv2}
	\end{equation}

For the case $n=0$ these solutions are reduced to ones for TEGR.

Also, we note, that the non-minimal coupling (for $n\neq0$) doesn't affect the shape of the scalar field potential (\ref{Sv}) compared to the potential for the TEGR case ($n=0$).


Now, we will consider the correspondence of this cosmological model to observational constraints on the values of the parameters of cosmological perturbations.


For Hubble parameter (\ref{S3}) from (\ref{SR1})--(\ref{SR2}) we obtain
\begin{eqnarray}
\label{SR3}
&&\epsilon=\frac{1}{(1+n)t}\left(\frac{m}{t}\right)^{-\frac{1}{1+n}},\\
\label{SR4}
&&\delta=\left(1+\frac{n}{2}\right)\epsilon.
\end{eqnarray}

Also, from (\ref{S3}), (\ref{PERTG2}) and (\ref{SR3}) we obtain
	\begin{equation}
		{\mathcal P}_{S}(t=t_{\ast})=\frac{\lambda^{n}t_{\ast}}{8\pi^{2}}\left(\frac{m}{t_{\ast}}\right)^{\frac{3}{1+n}},
		\label{PST}
	\end{equation}
where $t_{\ast}$ is the time of crossing of the Hubble radius.

Thus, we have following dependence the time of crossing of the Hubble radius for the model's parameters
	\begin{equation}
t_{\ast}=m\left(\frac{8\pi^{2}}{m\lambda^{n}}\times2.1\times10^{-9}\right)^{\frac{n+1}{n-2}}=
\frac{\lambda^{n}A^{2}}{2}
\left(\frac{16\pi^{2}}{\lambda^{2n}A^{2}}\times2.1\times10^{-9}\right)^{\frac{n+1}{n-2}},
		\label{PST1}
	\end{equation}
due to verification of the model under consideration in accordance with observational constraints.

Also, we consider the e-folds number
	\begin{equation}
N=\int^{t\ast}_{t_{0}} H(t)dt=\int^{t\ast}_{\mu t_{\ast}} H(t)dt,
		\label{N}
	\end{equation}
where $t_{0}$ is the time of beginning of inflation and $\mu=t_{0}/t_{\ast}$, also $0<\mu<1$.

Thus, from (\ref{PST1}) and (\ref{N}) we have
	\begin{equation}
N=\lambda^{n}A^{2}\left(\frac{1}{2}-\frac{\mu}{2}\right)\left(1+\frac{1}{n}\right)
\left(\frac{16\pi^{2}}{\lambda^{2n}A^{2}}\times2.1\times10^{-9}\right)^{\frac{n}{n-2}}\simeq60.
		\label{N1}
	\end{equation}

As one can see, from this condition it is impossible to unambiguously determine the parameters of the cosmological model, however, one can use this condition to estimate the remaining parameters of the model based on the given parameters.

Also, from (\ref{PERTG3})--(\ref{PERTG4}) and (\ref{SR4}) we obtain
\begin{equation}
r=8\left(\frac{1+n}{1-n}\right)(1-n_{S}),
\label{RNS}
\end{equation}
and the inflationary model under consideration corresponds to the observational constraints on the values of tensor-to-scalar ratio and spectral index of scalar perturbations for $-1<n<-0.77$.

\begin{figure}[ht]
\centering
\includegraphics[width=9 cm]{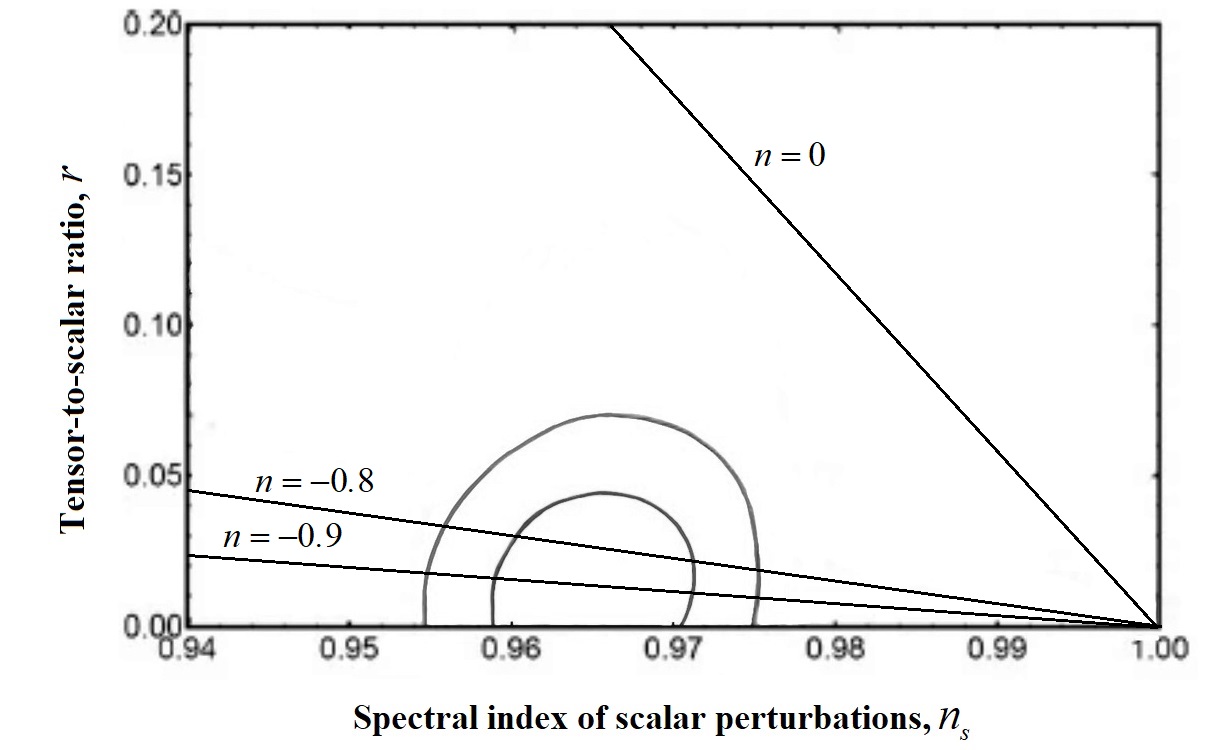}
\caption{The dependence $r=r(n_{S})$ for different values of the parameter $n=0,-0.8,-0.9$ with constraints on the tensor-to-scalar ratio $r$ due to the Planck TT,TE,EE+lowE+lensing+BAO+BICEP2/Keck Array  observations~\cite{Planck:2018vyg,Tristram:2021tvh}.}
\label{fig1PL2}
\end{figure}

On the Fig.\ref{fig1PL2} the dependence (\ref{RNS}) for the modified power-law inflation is represented.
Thus, for the case of minimal coupling ($n=0$) the model is not verifiable by observational constraints on the values of cosmological perturbations parameters, and for its verification it is necessary to take into account the non-minimal coupling of the scalar field and torsion.

Now we will consider the nature of cosmological dynamics for this model.

For this purpose, we define the relative acceleration of the expansion of the universe as follows
\begin{equation}
Q\equiv\frac{\ddot{a}}{a}=H^{2}+\dot{H}.
\label{PLQ}
\end{equation}

For the case of the Hubble parameter (\ref{S3}) we obtain
\begin{equation}
Q=\left(\frac{m}{t}\right)^{\frac{2}{1+n}}-\frac{\left(\frac{m}{t}\right)^{\frac{1}{1+n}}}{(1+n)t}.
\label{PLQ1}
\end{equation}

\begin{figure}[ht]
\centering
\includegraphics[width=9 cm]{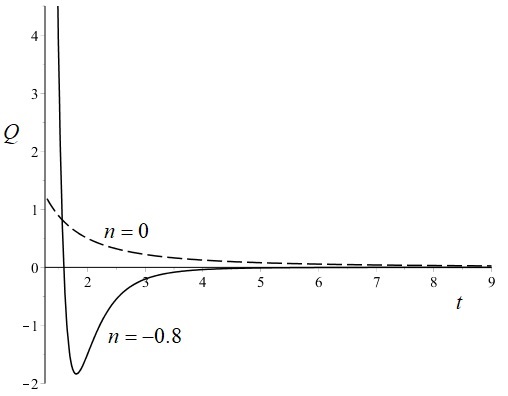}
\caption{The relative acceleration of the expansion of the universe for the Hubble parameter (\ref{S3}) for $n=0$ and $n=-0.8$.}
\label{fig1PL3}
\end{figure}

On the Fig.\ref{fig1PL3} the relative acceleration of the expansion of the universe for the cosmological model under consideration is shown.

As one can see, for the case of minimal coupling $n=0$ there is not exit from inflation $Q>0$ and at the large times $t\rightarrow\infty$ there is not accelerated expansion of the universe $Q=0$.

For the case of non-minimal coupling $n=-0.8$ there is exit from inflation, i.e. we have transition from stage with positive relative acceleration $Q>0$ to the stage (stages) with negative relative acceleration $Q<0$.
However, in this case there is also no stage of accelerated expansion of the universe at large times $Q=0$.

Thus, this model cannot be considered as a relevant one for describing the evolution of the universe.

\subsection{\label{sec:leve7}Modified exponential power-law inflation}

Now, we write corresponding Hubble parameter for the scalar field (\ref{S2}) for the case $C\neq0$ as follow
	\begin{equation}
		H(t)=\left(\beta+\frac{m}{t}\right)^{\frac{1}{n+1}},~~~~m=\frac{\lambda^{n}A^{2}}{2},
		\label{EPL}
	\end{equation}
where $\beta$ is an arbitrary constant.

Corresponding scale factor is can be defined in explicit form for the specific values of the parameter $n$ only.

From inverse dependence $t=t(\phi)$, connection between coupling function and the Hubble parameter $F=\left(H/\lambda\right)$ and Eq. (\ref{poten_4a}) we obtain
	\begin{equation}
		V(\phi)=\left\{\frac{3\beta}{\lambda^{n}}+\frac{3mB}{\lambda^{n}}e^{-\frac{\phi}{A}}\right\}
\left(\beta+mBe^{-\frac{\phi}{A}}\right)^{\frac{1}{n+1}}-\frac{mB^{2}}{\lambda^{n}}e^{-2\frac{\phi}{A}},
		\label{EPL1}
	\end{equation}
	\begin{equation}
		F(\phi)=\lambda^{-n}\left(\beta+mBe^{-\frac{\phi}{A}}\right)^{\frac{n}{n+1}}.
		\label{EPL2}
	\end{equation}

For the case $n=0$ these solutions are reduced to ones for TEGR.

Now, we consider the influence of non-minimal coupling between the  scalar field and torsion on the cosmological dynamics at the inflationary stage.

For this purpose we rewrite the Hubble parameter as follows
	\begin{equation}
		H(t)=\beta\left(1+\frac{m}{\beta^{n+1}t}\right)^{\frac{1}{n+1}}.
		\label{EPL4}
	\end{equation}

At the inflationary stage, the dynamics of the accelerated expansion of the universe is close to exponential one with the Hubble parameter $H\simeq const$, i. e. we have the condition $\frac{m}{\beta^{n+1}t}\ll1$.

Under this condition one can obtain expression for the Hubble parameter (\ref{EPL4}) as
	\begin{equation}
		H(t)\simeq\beta+\frac{\tilde{m}}{t},
		\label{EPL5}
	\end{equation}
where $\tilde{m}=\frac{m}{\beta^{n}(n+1)}=\frac{\lambda^{n}A^{2}}{2\beta^{n}(n+1)}$.

Corresponding scale factor
	\begin{equation}
		a(t)\propto t^{\tilde{m}}\exp\left(\beta t\right),
		\label{EPL5A}
	\end{equation}
coincides with the scale factor for TEGR.

Thus, in this case, the non-minimal coupling between the  scalar field and torsion does not change the type of accelerated expansion of the early universe.

From (\ref{PERTG2}) we have
	\begin{equation}
{\mathcal P}_{S}=\frac{\lambda^{n}\left(\beta+\frac{\tilde{m}}{t}\right)^{4}t^{2}}{8\pi^{2}(n+1)\tilde{m}}\simeq
\frac{\lambda^{n}\beta^{3}\left(\beta+4\frac{\tilde{m}}{t}\right)t^{2}}{8\pi^{2}(n+1)\tilde{m}}.	
		\label{EPL6}
	\end{equation}

Thus, the time of the crossing of the Hubble radius is
	\begin{equation}
t_{\ast}\simeq\left[\frac{8\pi^{2}\tilde{m}}{\lambda^{n}\beta^{4}}(n+1)\times2.1\times10^{-9}
+4\frac{\tilde{m}^{2}}{\beta^{2}}\right]^{1/2}-2\frac{\tilde{m}}{\beta}.	
		\label{EPL7}
	\end{equation}

From (\ref{N}) and (\ref{EPL5}) we have
	\begin{eqnarray}
\nonumber
&&N\simeq\beta t_{\ast}(1-\mu)-\tilde{m}\ln\mu\simeq\\
&&\simeq\beta(1-\mu)\left[\frac{8\pi^{2}\tilde{m}}{\lambda^{n}\beta^{4}}(n+1)\times2.1\times10^{-9}
+4\frac{\tilde{m}^{2}}{\beta^{2}}\right]^{1/2}+
\tilde{m}(-2+2\mu-\ln\mu)\simeq60.
		\label{N2}
	\end{eqnarray}

Also, for the Hubble parameter (\ref{EPL5}) from expression (\ref{SR3})--(\ref{SR4}) we obtain $\epsilon=\tilde{m}\delta^{2}$.

From expressions (\ref{PERTG3})--(\ref{PERTG4}) after neglecting the term of second order we obtain
	\begin{equation}
r=4\tilde{m}(n+1)(1-n_{S})^{2}.	
		\label{EPL8}
	\end{equation}

For $r<0.032$ and $n_{S}\simeq0.97$ we obtain the following condition on the model's parameters $\tilde{m}(n+1)=\frac{\lambda^{n}A^{2}}{2\beta^{n}}<9$.

\begin{figure}[ht]
\centering
\includegraphics[width=9 cm]{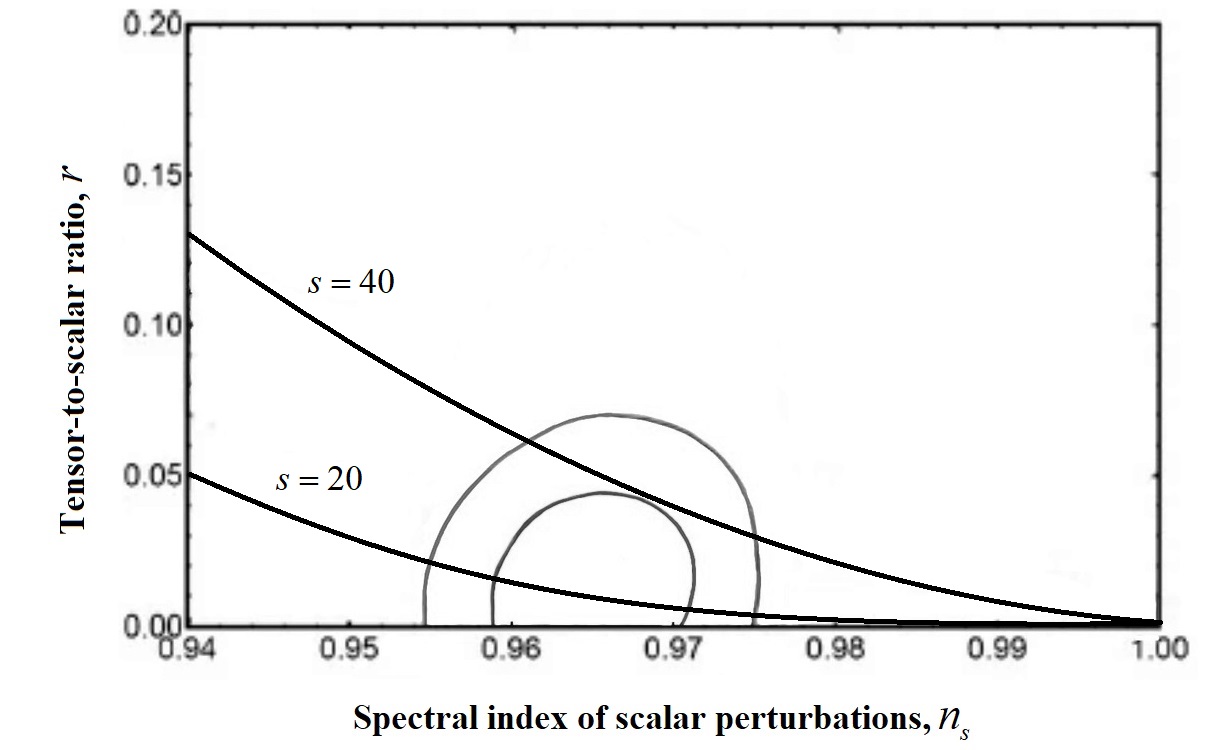}
\caption{The dependence $r=r(n_{S})$ for different values of the parameter $s=4\tilde{m}(n+1)=40,20$ with constraints on the tensor-to-scalar ratio $r$ due to the Planck TT,TE,EE+lowE+lensing+BAO+BICEP2/Keck Array  observations~\cite{Planck:2018vyg,Tristram:2021tvh}.}
\label{fig1PL4}
\end{figure}

On the Fig.\ref{fig1PL4} the dependence (\ref{RNS}) for the modified exponential power-law inflation is represented.
This type of inflation is verifiable for the both cases: minimal coupling ($n=0$) and non-minimal coupling ($n\neq0$) of the scalar field and torsion.

The relative acceleration (\ref{PLQ}) corresponding to the Hubble parameter (\ref{EPL}) is  
\begin{equation}
Q=\left(\beta+\frac{m}{t}\right)^{\frac{2}{n+1}}
+\frac{m\left(\beta+\frac{m}{t}\right)^{-\frac{n}{n+1}}}{(1+n)t^{2}}.	
\label{EPL8A}
\end{equation}

\begin{figure}[ht]
\centering
\includegraphics[width=9 cm]{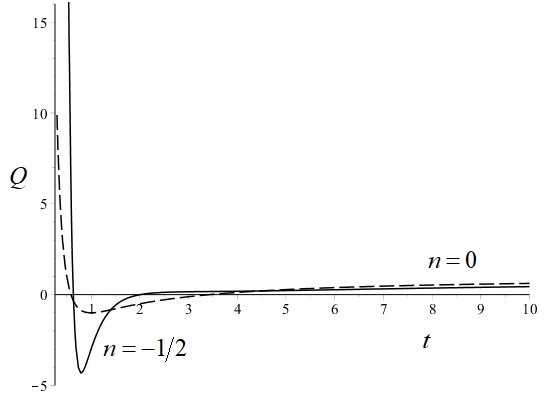}
\caption{The relative acceleration of the expansion of the universe for the Hubble parameter (\ref{EPL}) for $\beta=-1$ and $n=0,-1/2$.}
\label{fig1PL5}
\end{figure}

On the Fig.\ref{fig1PL5} the relative acceleration of the expansion of the universe for the cosmological model under consideration is represented.

For the case $\beta<0$ for the minimal coupling ($n=0$) and non-minimal coupling ($n\neq$) there are exits form inflation (from stage with $Q>0$ to the stage $Q<0$) and transitions to the second accelerated expansion of the universe (from stage with $Q<0$ to the stage $Q>0$).

Also, for $\beta<0$ from expression for the Hubble parameter (\ref{EPL}) we obtain condition of the constant parameter $n$ in the following form $\frac{1}{1+n}=2k$, where $k=1,2,3,...$  

The relative acceleration of the universe's expansion at the large times $t\rightarrow\infty$ is $Q=\beta^{4k}>0$, where $k=1/2$ corresponds to the minimal coupling.

Thus, the exponential power-law inflation can be considered as relevant cosmological models for TERG and for scalar-torsion gravity under condition (\ref{PLPM}).

\section{Conclusion}\label{Section5}

We have considered the influence of the non-minimal coupling of the scalar field and torsion on the cosmological dynamics and the potential of the scalar field for the case of a parametric connection $F\sim H^{n}$ between the non-minimal coupling function and the Hubble parameter.

A non-minimal coupling between the scalar field and torsion can induce significant changes in cosmological dynamics, as was shown in the example of power-law inflation. 

Also, for the power-law parameterization of the influence of modifications of the teleparallel equivalent of general relativity, it was shown that the potential of the scalar field changes as follows $V\simeq F(\phi)V_{(TEGR)}$.

It should be noted that taking into account the non-minimal connection makes it possible to verify models of cosmological inflation using observational constraints on the parameters of cosmological perturbations.

We also note that although the proposed approach to the construction and analysis of cosmological models was demonstrated for two cases of power-law and exponential-power inflation, it can be applied to arbitrary cosmological models.

As prospects for further research of cosmological models based on scalar-torsion gravity, we consider the analysis of models based on other physically motivated scalar field potentials and other dynamics of the accelerated expansion of the universe.

Another actual direction for further research is the calculation of the spectrum of relict gravitational waves for cosmological models based on scalar-torsion gravity and assessing the possibility of their direct registration.

\section*{Acknowledgements}

The study was supported by the Russian Science Foundation (Project No. 22-22-00248).

\end{document}